\documentclass[a4paper]{article}
\usepackage{multirow}
\usepackage{INTERSPEECH2019}
\usepackage{xcolor}
\usepackage{amsmath}
\usepackage{url}

\title{STC Antispoofing Systems for the ASVspoof2019 Challenge}
\name{Galina Lavrentyeva$^{1,2}$, Sergey Novoselov$^2$, Andzhukaev Tseren$^1$, Marina Volkova$^1$, Artem Gorlanov$^1$, Alexandr Kozlov$^1$}
\address{
  $^1$STC-innovations Ltd., St.Petersburg, Russia\\
  $^2$ITMO University, St.Petersburg, Russia \\ 
www.ifmo.ru \\
www.speechpro.com
}
\email{\{lavrentyeva, novoselov, andzhukaev, volkova, gorlanov, kozlov-a\}@speechpro.com}

\begin{document}

\maketitle
\begin{abstract}
This paper describes the Speech Technology Center (STC) anti-spoofing systems submitted to the ASVspoof 2019 challenge \footnote{This work was partially financially supported by the Government of the Russian Federation (Grant 08-08).}.
The ASVspoof2019 is the extended version of the previous challenges and includes 2 evaluation conditions: logical access use-case scenario with speech synthesis and voice conversion attack types and physical access use-case scenario with replay attacks. 
During the challenge we developed anti-spoofing solutions for both scenarios.
The proposed systems are implemented using deep learning approach and are based on different types of acoustic features.
We enhanced Light CNN architecture previously considered by the authors for replay attacks detection and which performed high spoofing detection quality during the ASVspoof2017 challenge.
In particular here we investigate the efficiency of angular margin based softmax activation for training robust deep Light CNN classifier to solve the mentioned-above tasks. 
Submitted systems achieved EER of 1.86\% in logical access scenario and 0.54\% in physical access scenario on the evaluation part of the Challenge corpora.
High performance obtained for the unknown types of spoofing attacks demonstrates the stability of the offered approach in both evaluation conditions.
\end{abstract}

\section{Introduction}
Over the past few years, voice biometric technologies have reached impressive performance, which can be confirmed by the results of the NIST Speaker Recognition Evaluation (SRE) Challenges \cite{NIST}.
Automatic Speaker Verification (ASV) systems are already used in security systems of socially significant institutions, in immigration control, forensic laboratories and for identity verification in Internet banking, and other electronic commerce systems. 

Alongside the increasing performance and increasing confidence in speaker recognition methods, the privacy level of the information with the necessity to protect it also increases. This leads to higher requirements for the reliability of the biometric systems including their robustness against malicious attacks.
Active fraudster attempts to falsify voice characteristics in order to gain unauthorised access referred to as spoofing attacks or presentation attacks (ISO/IEC 30107-1) are the biggest threat for voice biometric systems.
The widespread use of ASV systems and new approaches in machine learning has forced the significant quality improvement of these attacks. Many studies show that despite the high performance of the state-of-the-art ASV systems they are still vulnerable and the need in reliable spoofing detection methods for ASV systems is apparent.

Automatic Speaker Verification Spoofing and Countermeasures initiative (ASVspoof) 
has attracted the high interest of the research community to the task of unforeseen spoofing trials detection.
It has significantly pushed forward the development of spoofing detection methods by organizing ASVspoof Challenges in 2015 and 2017, that were aimed to develop countermeasures to detect speech synthesis with voice conversion attacks and replay attacks, respectively.

In 2019, the competition was held for the third time and was the extended version of the previous ones \cite{ASVspoof2019}. The task was to design the generalised countermeasures in 2 evaluation conditions: logical access use-case scenario with speech synthesis and voice conversion attack types and physical access use-case scenario with replay attacks.

For both scenarios, we proposed several systems based on the enhanced Light CNN architecture, considered by the authors for replay attacks detection in \cite{ASVspoof2017} and outperformed other proposed systems during ASVspoof2017 challenge.
The proposed systems are based on different types of acoustic features. 

This paper explores angular margin based softmax and batch normalization techniques for anti-spoofing systems quality improvements. 

Section \ref{sec:LCNN} describes the proposed modifications of the original LCNN-system for spoofing detection from \cite{ASVspoof2017} in details. Section \ref{sec:systems} contains the overview of all proposed single and submitted systems, while in section \ref{sec:res} the results obtained for these systems on the development and evaluation parts are presented and analysed.

It is worth mentioning that according to the evaluation plan all data used for training and evaluation was modelled using acoustic replay simulation. On the one hand, this helps to carefully control acoustic and replay configurations, but on the other hand, results raise some doubts about the usability of the considered systems for real-case scenarios. According to our experiments performed for spoofing attacks in real and emulated telephone channel \cite{phonespoof} systems trained for emulated conditions cannot detect spoofing attacks in real cases.



\section{LCNN system modifications}
\label{sec:LCNN}
All of the proposed systems for both scenarios were based on the enhanced Light CNN architecture previously used for replay attack detection \cite{ASVspoof2017}. 
The specific characteristic of Light CNN architecture \cite{lightcnn} is the usage of the Max-Feature-Map activation (MFM) which is based on Max-Out activation function \cite{goodfellow2013maxout}. Neural network with MFM is capable to choosing features which are essential for task solving. According to impressive results obtained by the authors in \cite{ASVspoof2017} for replay attacks, such type of networks can be successfully implemented for anti-spoofing.

\subsection{Front-End}
We explored several types of acoustic features as input for LCNN, all of them were used in a raw format.

Our experience in spoofing detection confirms that power spectrum contains useful information related to the speech signal and artifacts specific to different spoofing attacks and can be used as informative time-frequency representation for spoofing detection task. We used raw log power magnitude spectrum computed from the signal as features.
For this purpose, the spectrum was extracted via:
\begin{itemize}
    \item constant Q transform (CQT) \cite{CQT}
    \item Fast Fourier Transform (FFT)
    \item Discrete Cosine Transform (DCT)
\end{itemize}

Additionally, we considered cepstral coefficients from baseline systems, proposed by the organisers of the ASVspoof2019: Linear Frequency Cepstral Coefficients (LFCC) \cite{lfcc} obtained by the use of triangular filters in linear space for local integration of the power spectrum and Constant Q Cepstral Coefficients based on the geometrically spaced filters \cite{CQT} .
We explored efficiency of using simple enegry based Speech Activity Detector (SAD) for solving spoofing detection task for both PA and LA attack types.

\subsection{LCNN classifier}
In contrast to our LCNN system presented in \cite{ASVspoof2017} for replay attacks detection, the proposed systems are used not as high-level features extractor, followed by GMM scoring model. Instead of that LCNN was used here for final score estimation based on the low-level acoustic features. 

Additional steps of batch normalization were also used after MaxPooling layers to increase stability and convergence speed during the training process. 
The detailed architecture is described in Table \ref{tab:cnn}.

\begin{table}[th]
  \caption{LCNN architecture}
  \label{tab:cnn}
  \centering
  \resizebox{\columnwidth}{!}{
  \begin{tabular}{l l l r}
    \toprule
    \textbf{Type} & \textbf{Filter / Stride} & \textbf{Output} & \textbf{Params}   \\
    \midrule
    Conv\_1   & $5 \times 5$ / $1 \times 1$ & $863 \times 600 \times 64$ & 1.6K      \\
    MFM\_2    & $-$                         & $864 \times 600 \times 32$ & $-$      \\
    \midrule
    MaxPool\_3   & $2 \times 2$ / $2 \times 2$ & $431 \times 300 \times 32$ & $-$      \\
    \midrule
    Conv\_4     & $1 \times 1$ / $1 \times 1$ & $431 \times 300 \times 64$ & 2.1K      \\
    MFM\_5   & $-$                         & $431 \times 300 \times 32$ & $-$      \\
    BatchNorm\_6  & $-$                         & $431 \times 300 \times 32$ & $-$      \\
    Conv\_7     & $3 \times 3$ / $1 \times 1$ & $431 \times 300 \times 96$ & 27.7K     \\
    MFM\_8   & $-$                         & $431 \times 300 \times 48$ & $-$      \\
    \midrule
    MaxPool\_9   & $2 \times 2$ / $2 \times 2$ & $215 \times 150 \times 48$ & $-$      \\
    BatchNorm\_10  & $-$                         & $215 \times 150 \times 48$ & $-$      \\
    \midrule
    Conv\_11     & $1 \times 1$ / $1 \times 1$ & $215 \times 150 \times 96$ & 4.7K     \\
    MFM\_12   & $-$                         & $215 \times 150 \times 48$ & $-$      \\
    BatchNorm\_13  & $-$                         & $215 \times 150 \times 48$ & $-$      \\
    Conv\_14     & $3 \times 3$ / $1 \times 1$ & $215 \times 150 \times 128$ & 55.4K    \\
    MFM\_15   & $-$                         & $215 \times 150 \times 64$ & $-$      \\
    \midrule
    MaxPool\_16   & $2 \times 2$ / $2 \times 2$ & $107 \times 75  \times 64$ & $-$      \\
    \midrule
    Conv\_17    & $1 \times 1$ / $1 \times 1$ & $107 \times 75  \times 128$ & 8.3K     \\
    MFM\_18   & $-$                         & $107 \times 75  \times 64$ & $-$      \\
    BatchNorm\_19  & $-$                         & $107 \times 75 \times 64$ & $-$      \\
    Conv\_20     & $3 \times 3$ / $1 \times 1$ & $107 \times 75  \times 64$ & 36.9K     \\
    MFM\_21   & $-$                         & $107 \times 75  \times 32$ & $-$      \\
    BatchNorm\_22  & $-$                         & $107 \times 75 \times 32$ & $-$      \\
    
    Conv\_23    & $1 \times 1$ / $1 \times 1$ & $107 \times 75  \times 64$ & 2.1K     \\
    MFM\_24   & $-$                         & $107 \times 75  \times 32$ & $-$      \\
    BatchNorm\_25  & $-$                         & $107 \times 75 \times 32$ & $-$      \\
    Conv\_26     & $3 \times 3$ / $1 \times 1$ & $107 \times 75  \times 64$ & 18.5K     \\
    MFM\_27   & $-$                         & $107 \times 75  \times 32$ & $-$      \\
    \midrule
    MaxPool\_28   & $2 \times 2$ / $2 \times 2$ & $53 \times 37  \times 32$  & $-$      \\
    \midrule
    FC\_29       & $-$                     & $160$              & 10.2 MM     \\
    MFM\_30    & $-$                         & $80$                       & $-$      \\
    BatchNorm\_31  & $-$                       & $80$ & $-$      \\
    \midrule
    FC\_32         & $-$                   & $2$                        & 64      \\
    \midrule
    Total       & $-$                        & $-$                        & 371K     \\
    \bottomrule
\end{tabular}}
\end{table} 

\subsection{Angular margin based softmax activation}
The key difference of the novel LCNN system is angular margin based softmax loss (A-softmax) used for training the described architecture.
A-softmax was introduced in \cite{asoftmax} and demonstrated an elegant way to obtain well-regularized loss function by forcing learned features to be discriminative on a hypersphere manifold. Thus angular margin softmax loss can be described as:

\begin{equation}
\label{eq:asoftmaxloss}
\resizebox{.98\hsize}{!}{$%
L_\text{ang}=  \dfrac{1}{N}\sum_{i}-\log\left(\dfrac{e^{\|x_i\|\cos{(m\theta_{i,y_i})}}}{e^{\|x_i\|\cos{(m\theta_{i,y_i})}}+\sum_{i\neq y_i}e^{\|x_i\|\cos{(m\theta_{i,y_i})}}}\right)
$%
}
\end{equation}

where  $N$ is the number of training samples $\{x_i\}_{i=1}^N$ and their labels $\{y_i\}_{i=1}^N$, $\theta_{i,y_i}$ is the angle between $x_i$ and the corresponding column $y_i$ of the fully connected classification layer weights $W$, and $m$ is an integer that controls the size of an angular margin between classes. 

This approach has already used in \cite{odyssey2018} for high-level speaker embedding extractor. The learned features are constrained to a unit hypersphere. Such regularization technique also addresses the problem of overfitting by separating classes in cosine similarity metric.

We use A-softmax as an effective discriminative objective for training our model.

LCNN weights were initialized using normal Kaiming initialization. And dropout 0.75 was used to reduce overfitting.

\section{Experimental setup}
\subsection{Datasets}
All experiments presented further were conducted on ASVspoof 2019 datasets. The detailed description of these datasets can be found in \cite{ASVspoof2019}.
To train all the systems we used only the train part. The dev part was used for performance validation and weights adjustment for system fusion.
The evaluation part includes a set of unseen genuine verification trials and spoofing attacks, generated with unknown spoofing algorithms and replay configurations which differ from those in the train and development parts. 

\begin{figure*}[h]
	\centering
	\begin{minipage}[b]{0.23\textwidth}
		\includegraphics[trim={0.2cm 0 2.3cm 0}, width=\textwidth]{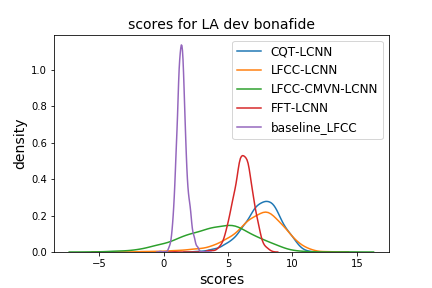}
	\end{minipage}
	\hfill
	\begin{minipage}[b]{0.23\textwidth}
		\includegraphics[trim={0.2cm 0 2.3cm 0}, width=\textwidth]{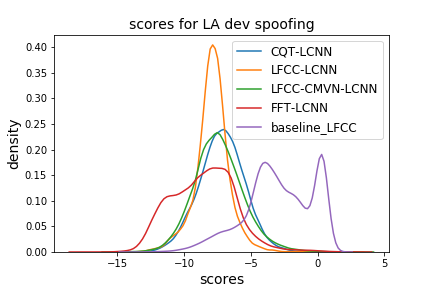}
	\end{minipage}
	\hfill
	\begin{minipage}[b]{0.23\textwidth}
		\includegraphics[trim={0.2cm 0 2.3cm 0}, width=\textwidth]{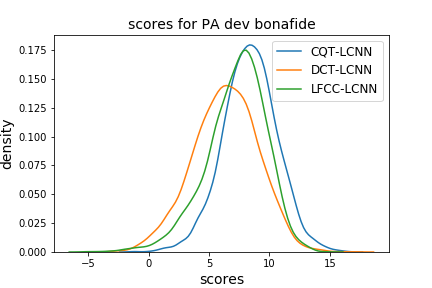}
	\end{minipage}
	\hfill
	\begin{minipage}[b]{0.23\textwidth}
		\includegraphics[trim={0.2cm 0 2.3cm 0}, width=\textwidth]{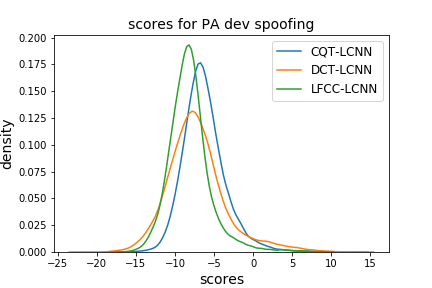}
	\end{minipage}
	\caption{Scores distributions for LA and PA systems for genuine and spoofing samples respectively.}
	\label{fig:distr}
\end{figure*}
\subsection{Details of systems implementation}
We prepared several single systems for each scenario, based on the features described above and LCNN architecture from \ref{tab:cnn}.
For logical access scenario we used the following configurations:
\begin{itemize}
\item \textbf{LFCC-LCNN}: LFCC were extracted similar to baseline system with 20 ms window length, 512 number of FFT bins and 20 filters.
\item \textbf{LFCC-CMVN-LCNN}: This system is similar to previous one. The only difference is that LFCC features were normalized by mean and variance.
\item \textbf{CQT-LCNN}: CQT spectrum was extracted with the use of the default settings from baseline CQCC based system: 96 bins per octave, 1724 window size and 0.0081 step.
\item \textbf{FFT-LCNN}: FFT spectrum was extracted with 1724 window length and step 0.0081, the Blackman window function was used. 
\end{itemize}
For physical access scenario we used the following configurations:
\begin{itemize}
    \item \textbf{LFCC-LCNN}: similar to LFCC-LCNN for LA scenario
    \item \textbf{CQT-LCNN}: similar to CQT-LCNN for LA scenario
    \item \textbf{DCT-LCNN}: For DCT spectrum the 863 window length and 0.0081 step were used.
\end{itemize}
Only the first 600 features for each file were used as LCNN input in all single systems. No additionally preprocessing techniques such as speech activity detection or dereverberation was explored in these systems.

\begin{figure}[h]
  \centering
  \begin{minipage}[b]{0.7\columnwidth}
    \includegraphics[width=\columnwidth]{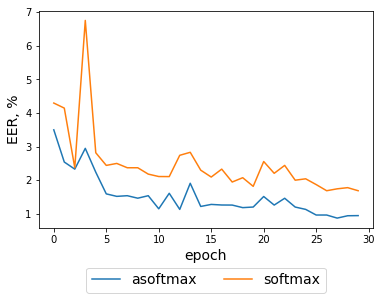}
  \end{minipage}
  \hfill
  \begin{minipage}[b]{0.7\columnwidth}
     \includegraphics[width=\textwidth]{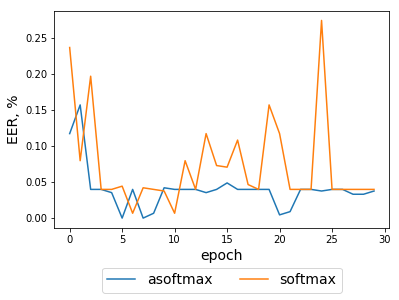}
  \end{minipage}
  \caption{EER during training process for PA CQT-LCNN system (top) and LA FFT-LCNN system (bottom)}
  \label{fig:train}
\end{figure}

\subsection{Submission systems}
\label{sec:systems}
The primary systems submitted to the challenge were the fusion of the single systems on the score level. Fusion of the subsystems scores was done with equal weights. Before fusion scores were normalized by the standard deviation of the genuine class distribution for each single system. The reason for that was the Gaussian distribution of genuine scores in contrast to spoofing scores, (see Figure \ref{fig:distr} for LA and PA systems scores distributions).

\section{Results and Discussion}
The results for our single and fusion systems are presented in terms of Equal Error Rate (EER) and minimum tandem detection cost function (min-tDCF) used as the primary metric in the Challenge.

Results obtained on the development and evaluation sets of the ASVspoof2019 dataset confirm the efficiency of deep learning approaches for the ASV spoofing detection tasks considered in the ASVSpoof2019.


\begin{table}[]
\centering
\caption{Performance of baseline systems and their modifications}
\label{tab:baseline}
\resizebox{\columnwidth}{!}{
\begin{tabular}{|l|l|l|l|l|}
\hline
              & \multicolumn{2}{c|}{LA} & \multicolumn{2}{c|}{PA} \\ \hline
System        & EER     & min-tDCF      & EER  & min-tDCF  \\ \hline
LFCC-GMM      &    3.029 &	0.078   &   11.226 &	0.241      \\ \hline
LFCC-CMVN-GMM &  6.000 & 0.153	      &    16.686 &	0.345        \\ \hline
LFCC-VAD-GMM  &   7.181 &	0.185 & 15.503 &	0.337         \\ \hline
CQCC-GMM      &   0.473 &	0.014	&	10.072 &	0.194         \\ \hline
CQCC-CMVN-GMM &  3.095	& 0.086	&	13.000 &	0.267         \\ \hline
CQCC-VAD-GMM &  3.571 &	0.108	&	10.144 &	0.204	       \\ \hline
\end{tabular}}
\end{table}

Results of the preliminary investigations of the baseline systems \cite{ASVspoof2019}, presented in Table~\ref{tab:baseline}, demonstrate that the use of SAD leads to the quality reduction in terms of EER and min-tDCF for LFCC and CQCC based systems in both LA and PA scenarios. The possible reason for this is that nonspeech and boundary regions contain discriminative features and distortions specific to various types of spoofing or genuine speech in the opposite. For example, constant energy values in concrete frequency regions, specific to some microphones or recording systems.
For this reason, we decided to exclude SAD from the systems we used in the Challenge.

\begin{figure*}[h]
	\centering \includegraphics[width=0.95\textwidth,trim=0 0.6cm 0 0.5cm]{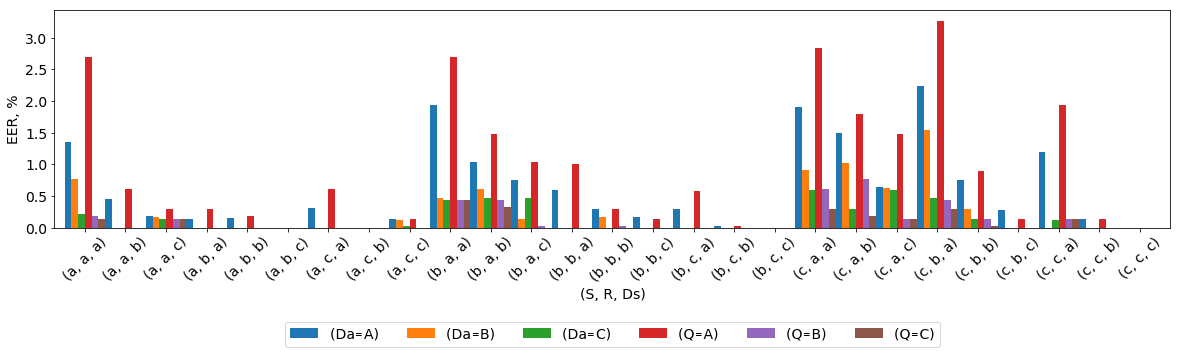}
	\caption{Performance of the primary PA system pooled by PA spoofing attack types from the evaluation set. $D_a$ relates to distance to a talker at which the replay attack is recorded, $Q$ reletes to loudspeaker quality, $S, R, D_s$ relates to (room size, reverberation and talker to ASv system distance)}
	\label{fig:pa_details}
\end{figure*}

Curious, that cepstral mean and variance (cmvn) features normalisation didn't provide an expected quality improvement on the development set (See Table \ref{tab:baseline}). 
This behaviour differs from the earlier experiments on ASVspoof2015 \cite{ASVspoof2015} and  ASVspoof2017 \cite{ASVspoof2017orig} datasets, that did not contain artificially produced data. \cite{ASVspoof2017upd}. 
Taking into account our experience in spoofing detection in unforeseen conditions we assume that cmvn can increase the robustness of our systems against unknown attacks from the evaluation set. However, according to results, of our single systems on the development and evaluation systems in Table \ref{tab:log_res}, we see the opposite.
Such results reinforce our concerns about modelled data and real case mismatch.

Experiment results for deep learning based systems, proposed in the current paper, prove that implementation of angular margin based softmax loss as classifier layer for spoofing detection system training allows to improve system quality and stabilize training process (see Figure \ref{fig:train}) for both LA and PA scenarios.

Experiments, conducted on the development part of ASVspoof2019 corpora, confirm that batch normalization and angular margin based softmax activation improve the performance of the original LCNN system for different types of low-level acoustic features in both scenarios (Figure~\ref{fig:train}).

Table \ref{tab:log_res} and Table \ref{tab:ph_res} present the performance of all single systems proposed for LA and PA respectively. High performance obtained for the unknown types of spoofing attacks performed on the evaluation part of ASVspoof2019 corpora demonstrates the stability of the offered approach in both evaluation conditions.
\begin{figure}[!h]
	\centering \includegraphics[width=\columnwidth,trim=0 0.5cm 0 1.0cm]{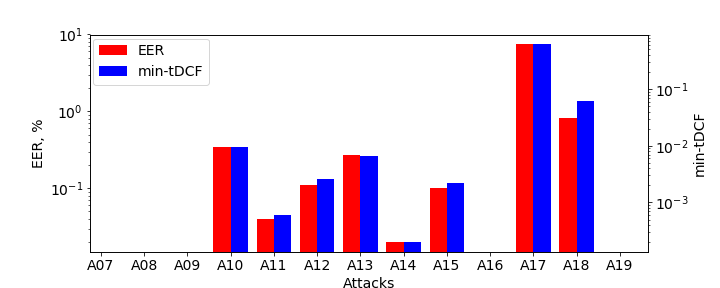}
	\caption{Performance of the primary LA system pooled by LA spoofing attack types from the evaluation set}
	\label{fig:LA_details}
\end{figure}

Detailed analysis of our LA final system quality for different types of logical attacks, that are presented in Figure \ref{fig:LA_details} demonstrates that it degrades in case of some unknown types of spoofing attacks (A10-A15, A17-A18) \cite{ASVspoof2019}.
The most difficult spoofing attack to detect for our system was A17 (voice conversion with waveform filtering) task for our system.

Figure \ref{fig:pa_details} illustrates the analysis of PA detection performance depended on the replay attack configuration: replay device quality, distances to the talker and to ASV system and reverberation characteristics. It can be concluded
that replay attack detection performance depends on the replay attacks quality. The most high-quality attacks replay sessions recorded at a small distance to talker with the use of high-quality loudspeaker.

\begin{table}[h]
\caption{Results for submitted LA systems}
\label{tab:log_res}
\centering
\resizebox{\columnwidth}{!}{
\begin{tabular}{|l|l|l|l|l|}
\hline
          & \multicolumn{2}{c|}{dev} & \multicolumn{2}{c|}{eval} \\ \hline
System    & min-tDCF     & EER       & min-tDCF      & EER       \\ \hline
LFCC-LCNN  & 0.0043      & 0.157      & \textbf{0.1000}        & 5.06      \\ \hline
LFCC-CMVN-LCNN & 0.0370  & 1.174     & 0.1827        & 7.86      \\ \hline
CQT-LCNN  & 0.0000       & 0.000     & -         & -      \\ \hline
FFT-LCNN  & 0.0009       & 0.040     & 0.1028         & \textbf{4.53}      \\ \hline
baseline\_LFCC      & 2.7060    & 0.069     & 0.2120         & 8.09      \\ \hline \hline
Fusion    &  0.0000            &  0.000         & \textbf{0.0510}        & \textbf{1.84}      \\ \hline
\end{tabular}}
\end{table}

\begin{table}[h]
\caption{Results for submitted PA systems}
\label{tab:ph_res}
\centering
\begin{tabular}{|l|l|l|l|l|}
\hline
          & \multicolumn{2}{c|}{dev} & \multicolumn{2}{c|}{eval} \\ \hline
System    & min-tDCF     & EER       & min-tDCF      & EER       \\ \hline
CQT-LCNN  & 0.0197       & 0.800      & \textbf{0.0295}        & \textbf{1.23}      \\ \hline
LFCC-LCNN & 0.0320       & 1.311     & 0.1053        & 4.60      \\ \hline
DCT-LCNN  & 0.0732       & 3.850     & 0.560         & 2.06      \\ \hline \hline
Fusion    &  0.0001  &  0.0154    & \textbf{0.0122}        & \textbf{0.54}      \\ \hline
\end{tabular}
\end{table}

\section{Conclusion}
This paper describes STC systems submitted to the ASVspoof2019 Challenge for LA and PA evaluation conditions. 
The main difference from the previous ASVspoof challenges is that all data used for training and  evaluation was modelled using acoustic replay simulation. In our opinion, this deals with some restrictions from the practical point of view.
According to the results obtained on the evaluation part of ASVspoof2019 corpora, the proposed LCNN based systems perform well in both PA and LA cases.
Submitted systems achieved EER of 1.86\% in LA scenario and 0.54\% in PA scenario for unknown types of attacks.

\bibliographystyle{IEEEtran}

\end{document}